# Mechanics, Energetics, Entropy and Kinetics of a Binary Mechanical Model System

Josh E. Baker[1*]

[1]Department of Pharmacology, University of Nevada, Reno School of Medicine, Reno, Nevada.

*Corresponding Author

**Email:**  jebaker@unr.edu

**Running Title: Thermodynamic Model of Force Generation**

**Abstract**

With the formal construction of a thermodynamic spring, I describe the mechanics, energetics, entropy, and kinetics of a binary mechanical model system. A protein that transitions between two metastable structural states behaves as a molecular switch, and an ensemble of molecular switches that displace compliant elements equilibrated with a system force constitutes a binary mechanical model system. In biological systems, many protein switches equilibrate with cellular forces, yet the statistical mechanical problem relevant to this system has remained unsolved. A binary mechanical model system establishes a limited number of macroscopic parameters into which structural and mechanistic details must be fit. Novel advances include a non-equilibrium kinetic and energetic equivalence; scalable limits on kinetics and energetics; and entropic effects on kinetics and mechanics. The model unifies disparate models of molecular motor mechanochemistry, accounts for the mechanical performance of muscle in both transient and steady states, and provides a new perspective on biomechanics with a focus here on how muscle and molecular motor ensembles work.

# Introduction

Many biomechanical mechanisms involve protein structural switches that generate or respond to an external force. For example, a conformational change in a protein (M) induced by ligand (A) binding has the potential to displace compliant elements and generate force upon binding (M to AM). Inversely, the force in compliant elements has the potential to reverse the conformational change and induce dissociation of the ligand (AM to M). A collection of molecular switches that generate force in thermally equilibrated compliant elements constitutes a binary mechanical model system. Binary model systems are often used to illustrate the basic principles of statistical mechanics with the typical example being that of a binary magnetic model system consisting of spins in a magnetic field; however, a binary mechanical model system applicable to an ensemble of protein switches remains undeveloped.

We previously solved the steady state case using a mean force field approach. With the formal construction of a thermodynamic spring, I have now solved the more general case. In a series of publications, I develop the model and its implications, showing that it reconciles disparate models of molecular motor mechanochemistry and accurately accounts for steady state and transient muscle mechanics and energetics. Here, I describe the energetics, mechanics, entropy, and kinetics of the system.

Springs are useful constructs for describing reversible changes in force and energetics in response to the displacement of molecules or molecular assemblies. They are "constructs" because they are not physically springs but rather representations of a physical relationship. They are "useful" only if they are accurate in this representation. In a thermodynamic system, molecular springs equilibrate with the system force and thus the mechanical state of a molecular spring within the system is thermally but not mechanically defined. The system force is the only mechanical constraint, and thus only one spring determines the relationship between force, energy, and displacement – a thermodynamic spring.

The key then to developing a binary mechanical model system is constructing a thermodynamically consistent system spring. Required of a binary model system, the spring must be constructed to describe an $F$-dependent Boltzmann distribution of states (M and AM). The spring must also be constructed to accurately describe changes in system force and energetics in response to displacements generated through both external changes in $F$ and internal switches. The structure of such a spring is described, and once constructed, a model of collective displacements of this spring by molecular switches is developed (a binary mechanical model). When the system is perturbed from equilibrium the system can generate force and, if allowed, contracts to perform work.

The model establishes a limited number of macroscopic parameters into which structural and mechanistic details must be fitted. It describes an equivalence between non-equilibrium kinetics and energetics that scales from single molecules to large ensembles. And it provides mechanistic descriptions of mechanochemical coupling, free energy transfer, thermodynamically consistent non-equilibrium rate constants, entropic effects on kinetics and mechanics, heat output, and both kinetic and energetic descriptions of contraction that when equated gives a chemical expression of Newton's third law (i.e., chemical and mechanical forces must be balanced during a contraction).

Muscle is a binary mechanical system within which the molecular switches are myosin motors (M), and the ligand that induces a conformational change (a lever arm rotation) is an actin filament (A) (1–7). The displacement of a compliant element by a switch-like myosin conformational change induced by actin binding is directly observed in single molecule mechanic studies (8–12). A switch-like conformational change upon actin binding is also observed in muscle where it has been shown to equilibrate with muscle force, $F$. Biochemical substates within these two mechanical states (13–15) are not considered in this analysis.

In 1938, little was known about actin and myosin; nevertheless, based on measurements of work and heat output by shorteninig muscle, A.V. Hill developed a thermodynamic model for muscle contraction that assumes that the chemistry of a force generating "contractile component" equilibrates with muscle force (16). Hill argued that this thermodynamic model was the framework into which the "detailed machinery must be fitted" (17). In contrast, molecular mechanic models, which provide the foundation for almost all muscle and molecular motors to date, require that thermodynamic parameters like the system force be fit into the detailed machinery (18, 19).

A binary mechanical model system is the detailed machinery fit into Hill's thermodynamic model. The development of this model was informed as much by mechanochemical studies of muscle as it was by the principles of thermodynamics, and the thermodynamic lessons learned provide new insights into systems biology. As the ideal binary mechanical model system, there is much left to be learned from muscle. Viewed through the lens of thermodynamics instead of molecular mechanics, a reanalysis of mechanochemical studies of muscle will provide fundamentally new perspectives on muscle and muscle systems. More broadly, a thermodynamic analysis of muscle provides an unparalleled window into how protein structure-function relationships in general scale up to cellular structure-function relationships under the influence of entropic forces, thermal fluctuations, and macroscopic constraints.

**Results**

**A Two State Scheme.** Figure 1 illustrates the two-state system. A protein (M) conformational change (a molecular switch) induced by ligand (A) binding displaces a compliant element a distance, $d$. The switch is reversible with forward, $f_+$, and reverse, $f_-$, rate constants defined below. Some springs irreversibly detach at a rate, $v$, through an active process that does not shorten the compliant element (Fig. 1).

**A Thermodynamic Spring.** Figure 2A (top) illustrates a network of $N_+$ parallel springs of stiffness, $\kappa_p$, each associated with an individual switch in series with a single spring of stiffness, $\kappa_s$. When the network of springs is held at a fixed length, a simple Hookean analysis shows that a single working step displaces the effective spring of stiffness, $\kappa_{sys}$, (Fig. 2A, bottom) from a length $F/\kappa_{sys}$ to a length $F/\kappa_{sys} + d/N_+$ performing reversible work

$$w_{rev} = \tfrac{1}{2}\kappa_{sys}\cdot(F/\kappa_{sys} + d/N_+)^2 - \tfrac{1}{2}\kappa_{sys}\cdot(F/\kappa_{sys})^2 = \tfrac{1}{2}\kappa_{sys}\cdot(d/N_+)^2 + F\cdot d/N_+,$$

which for $F > \tfrac{1}{2}\kappa_{sys}\cdot d/N$ is approximately

$$w_{rev} = F\cdot d/N_+ = \langle F\cdot d\rangle.$$

The F-dependence of the distribution of states is described by a Boltzmann distribution, $N_{AM}/N_M = \exp(-\langle F\cdot d\rangle)$, as required for a thermodynamic spring. Thermodynamic consistency is sufficient justification for constructing this particular spring; however, conceptuallly it describes compliant elements that are not localized to AM linkages and that are stretched regardless of the number of AM linkages that bear the load as long as there is at least one.

Not all compliant elements in the system are stretched, however. Forces must be balanced, and under certain conditions (e.g., during a contraction) a distribution of loads occurs with some compliant elements unstretched and others negatively stretched. Thus, the number, $N_+$, of switches that stretch compliant elements need not equal the total number of switches, $N$. For example, Fig. 2B shows that at a slack force a single switch ($N_+ = 1$) displaces a system spring a distance, $d$.

**Energetics of a Molecular Switch.** If the system spring rapidly equilibrates with a constant external force, $F$, the reaction free energy for binding is

$$\Delta_r G = \Delta G° + kT \cdot \ln(N_{AM}/N_M) + <F \cdot d>,  \qquad \text{Eq. 1}$$

where $\Delta G°$ is the standard free energy. Here we assume that the ligand concentration is fixed and is included in $\Delta G°$.

**Collective Displacement of the System Spring.** Force generation in the system spring is described by two master equations. A net flux of switches through the working step (Figure 1),

$$dN_{AM}/dt = -N_{AM} \cdot (f_- + v) + N_M \cdot (f_+), \qquad \text{Eq. 2}$$

collectively displaces the system spring a distance $d/N_+$ (Fig. 2A) generating force $\kappa_{sys} \cdot d/N_+$ in the system spring at a rate

$$dF/dt = [-N_{AM} \cdot (f_-) + N_M \cdot (f_+)] \kappa_{sys} \cdot d/N_+. \qquad \text{Eq. 3}$$

When $v = 0$, these equations describe the chemical relaxation of the binding reaction. The irreversible detachment of a switch ($v$ in Fig. 1) can occur through an active process that transfers a switch from AM to M ($v$ in Eq. 2) without a corresponding change in $F$ ($v$ does not appear in Eq. 3). This drives a net flux $N_M \cdot f_+ > N_{AM} \cdot f_-$ (Eq. 3) and completes a reaction cycle.

The energy difference between AM and M

$$\Delta G = \Delta G° + kT \cdot \ln(N_{AM}/N_M) + <F \cdot d>]$$

is added to the system when AM is irreversibly transferred to M. This process clearly requires an energy source, and for molecular motors, the source of energy is typically the free energy for ATP hydrolysis, $\Delta G_{ATP}$. Nevertheless, independent of $\Delta G_{ATP}$ the switch-ligand binding energy, $\Delta G$, is the energy available for work. The reversible binding performs work while irreversible detachment (driven by $\Delta G_{ATP}$) adds binding energy to the system. If the binding energy, $\Delta G$, exceeds $\Delta G_{ATP}$, then $\Delta G_{ATP}$ would be insufficient to detach a switch irreversibly, which is to say $\Delta G_{ATP}$ limits $\Delta G$.

**Stall Force.** Force is collectively generated by a net flux through the binding reaction until a "stall" force, $F_o$, is reached. This occurs when the free energy available for force generation (Eq. 1) approaches zero ($\Delta G = 0$) or, from Eq. 1, when the force per force-generating switch, $F/N_+$, is

$$F_o/N_+ = -[\Delta G^o + kT\ln(N_{AM}/N_M)]/d \qquad \text{Eq. 4}$$

Substituting back into Eq. 1, the energy available for work per force generating switch is

$$\Delta_r G = \langle F_o \cdot d \rangle - \langle F \cdot d \rangle,$$

and the free energy available for work by $N_+$ force-generating switches is

$$N_+ \cdot \Delta_r G = (F_o - F) \cdot d$$

**Work and Heat Output.** Expanding upon our previous analysis (20), in a steady state, when the system is held at a constant external force ($F < F_o$), switches generate forces (Fig. 3A, top) that rapidly equilibrate with the system force, $F$, when the system spring shortens a distance $x$ (Fig. 3A, bottom). Work and heat output are generated by shortening against both $F$ and a frictional, $F_f$, load (dashpot in Fig. 3 and resistive switches in Fig. 3B).

The energy available for work, $(F_o - F) \cdot d$, is irreversibly transferred to the surroundings as work, $F \cdot x$, and heat, $F_f \cdot x$, at the rate, $v$,

$$v \cdot (F_o - F) \cdot d = d(F \cdot x)/dt + d(F_f \cdot x)/dt,$$

or

$$v \cdot (F_o - F) \cdot d = F \cdot V + F_f \cdot V, \qquad \text{Eq. 5}$$

This is A.V. Hill's formalism into which a binary mechanical model has been fitted. To characterize the relationship between muscle force, $F$, and muscle shortening velocities, $V$, muscle physiologists universally use Hill's F-V equation:

$$b_H(F_o - F) = F \cdot V + a_H \cdot V.$$

This equation is widely considered to be phenomenological when, in fact, it is based on first principles (Eq. 5) with $b_H = v \cdot d$ and $a_H = F_f$.

**Unloaded Shortening, $V_o$.** According to Eq. 5, when $F = 0$, the maximum unloaded shortening velocity is energetically defined as

$$V_o = F_o \cdot v \cdot d / F_f.$$

The kinetic definition of $V_o$ is simply the product of the rate of stepping $N \cdot v$ and the apparent step size, $d/N_+$, or

$$V_o = N \cdot v \cdot d / N_+,$$

Comparing these energetic and kinetic definitions of $V_o$

$$N_+/N = F_f/F_o. \qquad \text{Eq. 6}$$

Written as $(N_+/N) \cdot F_o = F_f$, this equivalence of kinetics and energetics is simply a chemical version of Newton's third law. That is, the chemical force generated by $N_+$ switches is balanced by the frictional force, $F_f$. There are two limits to $V_o$. A minimum $V_o$ ($= v \cdot d$) is energetically limited by $F_f$

= $F_o$ ($F_f$ cannot exceed the maximum force generated, $F_o$), and a maximum $V_o$ (= $N \cdot v \cdot d$) is kinetically limited by $N_+ = 1$ (at least one switch must be force generating).

In muscle, $V_o$ exceeds the energetic limit, $v \cdot d$, by a factor $1/r$ of approximately 10 ($r = 0.1$) (21, 22). According to a molecular mechanic model, $V_o$ is energetically limited, and $r$ is defined as the fraction of myosin motors strongly bound to actin (21). In a thermodynamic model, $r$ is the extent, $F_f/F_o$, to which $V_o$ is energetically limited, which is also the thermodynamic duty ratio, $N_+/N$, and the curvature, $a_H/F_o$, of the F-V relationship with values ranging from 0.1 to 0.5 (8). Using an in vitro motility assay, we have demonstrated that $V_o$ is not energetically limited (12, 23–26) and recently estimated a value for $F_f/F_o$ in this assay of approximately 0.4 (26).

**$N_+$ Varies Linearly With $F$.** When $F > 0$, Eq. 6 becomes

$$N_+ = N \cdot [(F_f + F)/F_o], \qquad \text{Eq. 7}$$

which implies that the number of force-generating switches, $N_+$ (not the energetically limited maximum force generated per switch, $F_o/N$) changes with the applied force, $F$. The mechanism by which $N_+$ decreases with decreasing $F$ is that higher rates of shortening at lower forces redistribute compliant elements from positive to neutral and eventually to resistive forces. The likely V-dependence of $F_f$ is not considered in this analysis.

**System Entropy.** The entropy of the system is $k \cdot \ln\Omega$, where $\Omega$ is the number of available microstates. For a given $N_M$, there are $\Omega = N!/[N_M! \cdot (N - N_M)!]$ microstates. When one switch binds to a ligand, $N_M$ decreases by one, changing the system entropy by

$$\Delta S = k \cdot \ln(N!/[(N_M - 1)! \cdot (N - N_M + 1)!]) - k \cdot \ln(N!/[N_M! \cdot (N - N_M)!]),$$

which reduces to $\Delta S = -k \cdot \ln[(N_{AM} + 1)/N_M]$, and for large $N_{AM}$ is approximately

$$\Delta S = -k \cdot \ln[N_{AM}/N_M].$$

Thus, $kT \cdot \ln[N_{AM}/N_M]$ in Eq. 1 is the negative reaction entropy, $-T\Delta S$.

**Entropic Force.** According to Eq. 1, when both $\Delta_r G$ and $\Delta G^o$ equal zero, $F = -N_+ \cdot kT \cdot \ln(N_{AM}/N_M)/d$ (Fig. 4A), which according to the analysis above is an entropic force. During active force generation, $N_{AM}/N_M$ decreases with increasing $F$ and when the entropic energy $-kT \cdot \ln[N_{AM}/N_M]$ becomes positive, it contributes to a negative $\Delta G$ as energy available for work. However, $T\Delta S$ cannot be used for $F \cdot x$ work, which means that it must be lost from the system as heat. When force generation is non-adiabatic (heat is lost from the system, e.g., through movement against $F_f$), no entropic force is generated, $kT\ln(N_{AM}/N_M) = 0$, and a stall force, $F_o = -N_+ \cdot \Delta G^o/d$, is reached when $N_{AM} = N_M$.

However, if force generation is adiabatic (heat is not lost from the system, e.g., because no movement is allowed) the irreversible transfer of AM to M through $v$ generates entropic force, reaching a maximum limit when only one switch remains bound [$N_{AM}/N_M = 1/(N-1)$]. This occurs at a maximum entropic force of $F = -N_+ \cdot kT \cdot \ln(N-1)/d$ that scales with $N$ (entropic force is not defined for one molecule and the average entropic force per switch increases with $N$). Force generation beyond this entropic limit results in the detachment of the one remaining bound switch, dissipating the energy in the system spring as heat. We show elsewhere that the resulting periodic entropic force generation accounts for periodic force generation observed in certain myosin systems (27).

Figure 4B shows that entropic force is buffered by incremental displacements, $\Delta x$, in $x$, generating relatively small entropic forces when $N_{AM}/N_M$ is between 0.5 and 2 and larger forces when $N_{AM} \ll N_M$ or when $N_{AM} \gg N_M$. The slope, $\Delta F/\Delta x$, of the curve in Fig. 4B is an entropic stiffness, which is relatively low in the buffered range and increases at both ends of the concentration gradient.

**Non-Equilibrium Rate Constants ($f_+$ and $f_-$).** A non-equilibrium system has a net flux, $N_M \cdot f_+ > N_{AM} \cdot f_-$, which can be written

$$kT \cdot \ln(N_{AM}/N_M) - C = kT \cdot \ln(f_+/f_-),$$

where C is the energy that drives the forward reaction. The irreversible transfer of a switch from AM to M through $v$ in Fig. 1 adds $\Delta G = \Delta G° + kT \cdot \ln(N_{AM}/N_M) + <F \cdot d>$ to the system, which drives the net flux (C = $\Delta G$). When $\Delta G$ is negative, the reaction is spontaneous in the forward direction; when $\Delta G$ is positive, the reaction is spontaneous in the reverse direction. Thus,

$$kT \cdot \ln(N_{AM}/N_M) - \Delta G = kT \cdot \ln(f_+/f_-).$$

Substituting $kT \cdot \ln(N_{AM}/N_M)$ from Eq. 1,

$$\Delta_r G - \Delta G° - <F \cdot d> - \Delta G = kT \cdot \ln(f_+/f_-).$$

Here, the reaction free energy, $\Delta_r G$, and the free energy added, $\Delta G$, do not cancel. They are not opposing chemical forces that prevent a chemical relaxation by somehow bringing the non-equilibrium system to equilibrium. Neither do they spontaneously dissipate to bring the system to equilibrium without a chemical relaxation. Both $\Delta_r G$ and $\Delta G$ approach zero at equilibrium, but the relaxation reaction constrained by Eqs. 2 and 3 must occur to get there. Only the free energy, $\Delta G$, that is physically added to the system drives the relaxation reaction ($f_+/f_-$ is more positive when $\Delta G$ is more negative). The reaction free energy, $\Delta_r G$, describes a decrease in free energy as the reaction proceeds, which clearly does not drive the reaction ($f_+/f_-$ is more negative when $\Delta_r G$ is more negative). Solving for $f_+/f_-$

$$f_+/f_- = \exp[(-<F \cdot d> - \Delta G^\circ - \Delta G)/kT] \qquad \text{Eq. 8}$$

This equation describes the kinetics of both adiabatic (or entropic) and non-adiabatic force generation. When $\Delta G$ is freely exchanged with the surroundings (non-adiabatic), the chemical relaxation occurs toward $\Delta G = 0$, where at equilibrium $f_+/f_- = 1$ and $<F \cdot d> = -\Delta G^\circ$. This is consistent with the energetic argument above that a non-adiabatic equilibrium is reached when $N_{AM}/N_M = 1$ and $<F \cdot d> = -\Delta G^\circ$ (together $f_+/f_- = N_{AM}/N_M$, consistent with no net flux at equilibrium). When $\Delta G$ is not lost to the surroundings as heat (adiabatic), the chemical relaxation occurs toward $\Delta G < 0$, which drives the chemical relaxation past the non-adiabatic force, generating entropic force as described above. With increasing adiabatic force generation, $N_{AM}/N_M$ decreases, and $T\Delta S = -kT\ln(N_{AM}/N_M)$ increases which makes $\Delta G$ more negative, providing additional energy to energetically ($\Delta G < 0$) and kinetically (Eq. 8) generate additional entropic force.

We show elsewhere that adiabatic force generation followed by equilibration accounts for phases 2 and 3 of a muscle force transient. Each energy term, $E$, in Eq. 8 can be independently partitioned between $f_+$ and $f_-$ depending on the extent, $a_E$, to which each energy change occurs prior to the activation energy barrier for the working step. For example, at equilibrium ($\Delta G = 0$) and $F = 0$, standard solution rate constants are

$f_+^\circ = \exp[a_{Go} \cdot \Delta G^\circ / kT]$ and

$f_-^\circ = \exp[(1 - a_{Go}) \cdot \Delta G^\circ / kT]$

Figures 5A and 5B are simulations of force generation at $\Delta G^\circ = 0$ and different $\kappa_{sys}$ values. These simulations demonstrate that entropy ($-k \cdot \ln[N_{AM}/N_M]$) alone can generate force. Interestingly, Figs. 5A and 5B show that the rate of entropic force generation increases with the system spring stiffness, $\kappa_{sys}$. Panel insets in Fig. 5 are the corresponding simulated time courses of the energetic terms in Eq. 1 illustrating that these simulations are thermodynamically consistent; that is, simulated reactions intrinsically occur in a direction dictated by ΔG, and equilibrate at values for $F$, $N_{AM}$, and $N_M$ that satisfy Eq. 1 at ΔG = 0. This thermodynamic consistency is observed in all simulations regardless of initial conditions since rate constants are derived directly from the free energies (Eq. 1).

**Discussion**

It has taken over 80 years to fit the "detailed machinery" into A.V. Hill's thermodynamic model of muscle contraction. Key to this development was constructing a thermodynamic spring. Springs in biology are formal constructs for describing physical relationships between force, energetics, and displacements. A thermodynamic spring must, therefore, describe an $F$-dependent Boltzmann distribution of states. The thermodynamic spring in Fig. 2B satisfies this requirement in addition to describing the relationship between force, energetics, and displacements generated both externally and internally by molecular switches.

A binary mechanical model establishes a limited number of macroscopic parameters ($d$, $\Delta G^o$, $y$, $\kappa_{sys}$, $v$) into which all structural and mechanistic details must be fitted. Unfortunately, in the 1950s images of myosin motors protruding from myosin filaments and interacting with adjacent actin filaments (28, 29) inspired a bottom-up molecular mechanical determinism not afforded by thermodynamics (19). Thus, instead of fitting the detailed machinery into thermodynamic parameters, thermodynamic properties of the system were fit into molecules, adding states and mechanisms as needed to account for properties and processes such as thermal equilibration and system entropy not reasonably contained within a molecule. Gibbs referred to this as seeking "mechanical definitions of temperature and entropy" (30).

The central assumption of the molecular mechanic formalism – which continues to be the foundation for most models of muscle and molecular motor ensembles to date – is that the system force can be projected onto a molecule and then defined by a thermally isolated molecular spring (19). While this molecular spring may provide a useful construct for imagining molecular mechanisms of system mechanics, it is not physical and has proven to be a distraction from the remarkable insights that mechanochemical studies of muscle provide into the thermodynamics of muscle and other biological systems.

Informed by mechanochemical studies of muscle, the binary mechanical model system developed here provides a remarkable tool for understanding chemical thermodynamic forces. In this model, force in a thermodynamic spring serves as a formal probe of the chemical thermodynamic forces generated by a simple binding reaction. These same chemical forces are generated by a binding reaction in the absence of a thermodynamic spring and can be generated against any macroscopic force such as membrane potentials, concentration gradients, the cytoskeleton, pressure, surface tension, extra cellular matrix, etc. Particularly significant are definitions of entropic forces, thermodynamically consistent non-equilibrium kinetics, adiabatic kinetics and energetics, thermal equilibration with macroscopic constraints, scalable kinetics and energetics, and energetic and kinetic equivalence, which provide formal guidance for understanding how the structure-function of isolated proteins scale up to function in cells.

Muscle is an ideal binary mechanical model system that offers almost limitless intriguing mechanochemical problems. A binary mechanical model analysis will provide new perspectives on all of these problems, which in turn will provide additional insights into chemical thermodynamics in biological systems.

**Acknowledgments**: I thank Willard, AV, Julie and my students, colleagues, and mentors who have over many years inspired and guided this work. This work was funded by a grant from the National Institutes of Health 1R01HL090938-01.

**Figure 1**. A two-state scheme. A conformational change in a molecular switch (M) is induced upon ligand (A) binding, displacing compliant elements a distance, $d$, at a rate, $f_+$. The reverse transition occurs at a rate $f_-$. Switches are irreversibly detached at a rate $v$ through an active pathway that does not revese the displacement of compliant elements.

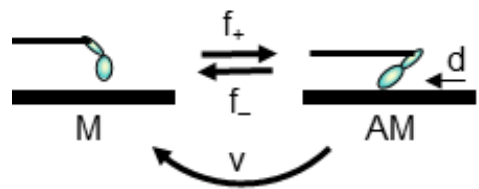

**Figure 2.** A single effective system spring held at a constant length (isometric). (A) A network of $N_+$ parallel springs with stiffness, $\kappa_p$, each associated with individual switches are all in series with a single spring of stiffness, $\kappa_s$ (top). When the network of springs is held at a fixed length, a simple Hookean analysis shows that the apparent distance a single switch displaces the series spring is $d/N_+$. Thus, the system spring can be defined (bottom) as a spring with stiffness, $\kappa_{sys}$ [$= 1/(1/\kappa_s + 1/N_+\kappa_p)$] that is stretched a distance $d/N_+$ (bottom) by a given switch. (B) When the system spring is held at a fixed length with $F = 0$, a single switch ($N_+ = 1$) displaces the system spring a distance, $d$ (assuming other switches do not interfere).

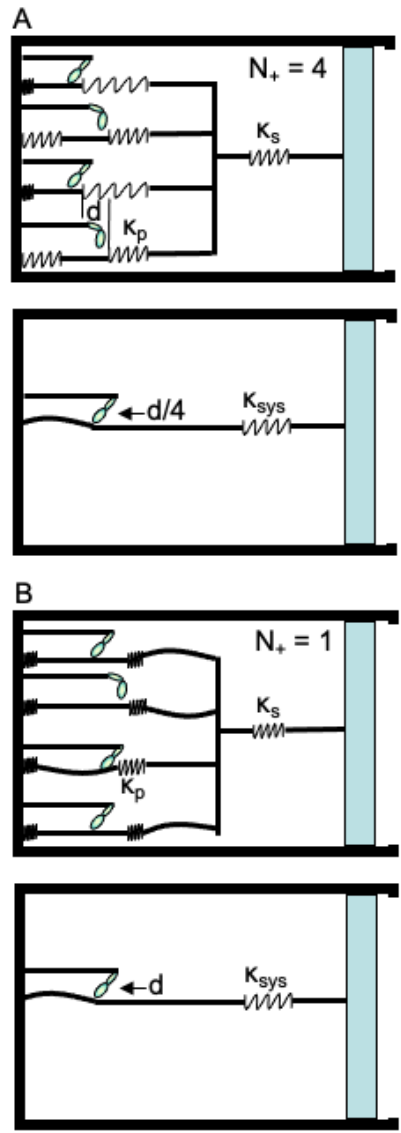

**Figure 3.** A single effective system spring held at a constant force, $F$ (isotonic). (A) When the system spring is held at a constant force, $F$, a switch displaces the system spring a distance, $d/N_+$, transiently generating force (top) that exceeds $F$. The system spring equilibrates with the constant force by moving a distance $x = d/N_+$ against $F$, performing $Fx$ work and losing heat through a viscous element (blue bar) to the surroundings. (B) As $F$ decreases, the number, $N_+$, of force generating switches decreases as switches are pulled through movement into relaxed or even resistive mechanical states, increasing both the apparent step size, $d/N_+$ and $F_f$ (resistive switches). The relationship between $N_+$ and $F_f$ is described by Eq. 6. At $F = 0$, force generation against $F_f$ (top) equilibrates with $F_f$ upon dissipative movement $x$ (bottom).

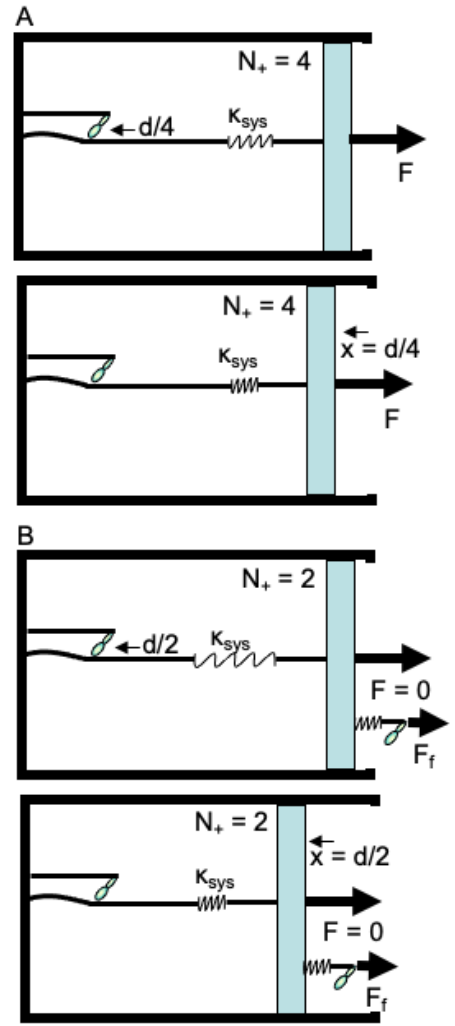

**Figure 4.** Relationships between $F$, $-\Delta G^\circ$, $N_{AM}$, and $N_M$ at equilibrium (Eq. 1, $\Delta_r G = 0$). (A) Equation 1 is plotted with $\Delta G^\circ = 0$ showing that the entropic force increases with $N_M$ (black squares), and the average force per bound switch (red circles) increases exponentially with $F$. (B) Using master equations 2 and 3, the system spring is stretched through 2 nm displacements (x axis) and the relaxation to a new equilibrium is simulated to obtain a new equilibrium force, $F$. Values for F are plotted at each 2 nm increment (x axis). The slope is entropic stiffness. At $\Delta x = 0$, initial values are $N_{AM} = 49$, $N_M = 1$, $\kappa_{sys} = 25.3$ pN/nm, $d = 12$ nm.

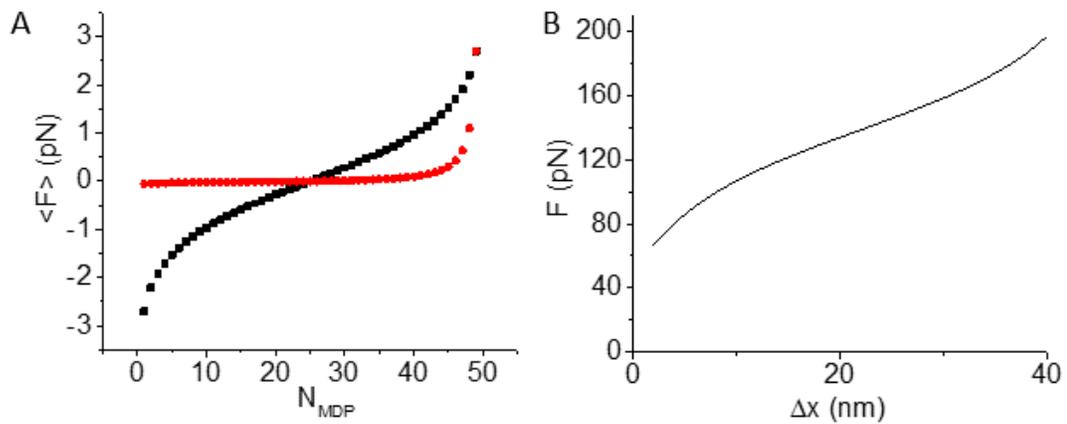

**Figure 5.** Chemical relaxations simulated using master equations 2 and 3. Time courses of $N_M$ (A) and $x_{eff} = F/\kappa_{sys}$ (B) are simulated with initial conditions $\Delta G° = 0$ ($f_+° = f_-° = 1$ sec$^{-1}$), $N_M = 49$ (blue lines) and $N_M = 30$ (black lines) at $\kappa_{sys} = 0$ (dashed lines) and $\kappa_{sys} = 5$ pN/nm (solid lines), $d = 12$ nm. (B, Inset) The time course of the energy terms in Eq. 1 are plotted for initial conditions $N_M = 30$ and $\kappa_{sys} = 5$ pN/nm.

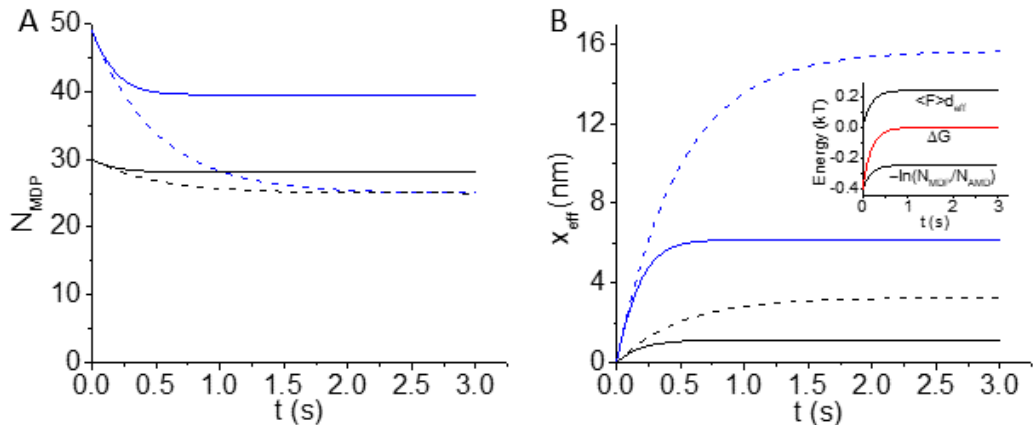